\documentclass{PoS}

\title{
\vspace{-7ex}\begin{minipage}[t][1pt][r]{\textwidth}
\begin{flushright}
\rm\small
FR-PHENO-2010-021
\end{flushright}
\end{minipage}\\[6ex]
NLO QCD corrections to 4 b-quark production}

\ShortTitle{NLO QCD corrections to 4 $b$-quark production}

\author{\speaker{Nicolas Greiner}\\
        University of Illinois at Urbana-Champaign, Urbana IL, 61801 USA\\
        E-mail: \email{ngreiner@illinois.edu}}

\author{Alberto Guffanti\\
        Physikalisches Institut, Albert-Ludwigs-Universit\"at Freiburg, 
        79104 Freiburg i. Br., Germany\\
        E-mail: \email{alberto.guffanti@physik.uni-freiburg.de}}

\author{Jean-Philippe Guillet\\
        LAPTH, Annecy-le-Vieux 74951, France\\
        E-mail: \email{guillet@lapp.in2p3.fr}}

\author{Thomas Reiter\\
        Nikhef, 1098XG Amsterdam, The Netherlands\\
        E-mail: \email{thomasr@nikhef.nl}}

\author{J\"urgen Reuter\\
        Physikalisches Institut, Albert-Ludwigs-Universit\"at Freiburg,
        79104 Freiburg i. Br, Germany\\
        E-mail: \email{juergen.reuter@physik.uni-freiburg.de}}

\abstract{In this talk we present the recent calculation of the NLO QCD corrections to the production
of four b-quarks induced by initial state quarks at the CERN LHC. We describe the details of the calculation
using the GOLEM approach for the virtual corrections and MadEvent/MadDipole for the real emission part and
present some results.}

\FullConference{XVIII International Workshop on Deep-Inelastic Scattering and Related Subjects, DIS 2010\\
		April 19-23, 2010\\
		Firenze, Italy}

\newcommand{\nn}{\nonumber}

\begin{document}

\section{Introduction}
The study of the Higgs sector of the Standard Model (SM) and the search for 
New Physics at the TeV scale at the LHC experiments at CERN will ultimately 
depend on a detailed understanding of signal and background processes involving
many particles \cite{:2008uu,Bern:2008ef,Buttar:2006zd,Campbell:2006wx}.

When considering hadronic collisions the results obtained using only leading
order (LO) in perturbation theory retain a strong dependence on the unphysical
factorisation and renormalisation scales which limits the predictivity of the
theoretical computations.


Whereas the inclusion of next-to-leading order (NLO) QCD effects is relatively 
straightforward for simple kinematic situations ($2\to 2$ and $2\to 3$ processes), processes 
with four parton final states are at the forefront of the present computing capabilities and the 
number of available results is very limited.
However a lot of progress has been made during the last couple of years.
Recent achievements are the evaluation of the order $\alpha_s$ corrections to 
$pp\to b\bar{b}t\bar{t}$ \cite{Bredenstein:2008zb,Bredenstein:2009aj,Bevilacqua:2009zn}, 
$pp\to Wjjj$ \cite{Berger:2009ep,Berger:2009zg,KeithEllis:2009bu,Ellis:2009zw},
$pp \to t \bar{t} jj$ \cite{Bevilacqua:2010ve} and $pp \to Z/\gamma jjj$ \cite{Berger:2010vm}.

In this contribution we present the results obtained in~\cite{Binoth:2009rv} for the process 
$pp \to b \bar{b} b \bar{b}$ which is a relevant background for Higgs searches in 
the minimal supersymmetric extension of the Standard Model (MSSM). The MSSM contains
two Higgs doublets containing five scalar Higgs bosons. As light Higgs bosons prefer to
decay into $b$-quarks in large parts of parameter space, experimental studies are so 
far strongly affected by the uncertainty of the Standard Model backgrounds, especially
$pp \to b\bar{b}b\bar{b}$, see for example Ref.~\cite{Lafaye:2000ec}. 
Although mainly motivated by supersymmetry, the four-$b$ final state also allows
the study of other interesting  scenarios for Beyond Standard Model (BSM) physics, 
such as Hidden Valley models, where decays of hadrons of an additional confining gauge group can produce 
high multiplicities of $b\bar{b}$ pairs~\cite{Bern:2008ef,Krolikowski:2008qa}.
This is the reason why the $pp\to b\bar{b}b\bar{b}$ process was added to the experimenters 
wishlist of relevant next-to-leading order computations \cite{Bern:2008ef}.

\section{Calculation}

At leading order two partonic subprocesses contribute to the computation of the $pp\to b\bar{b}b\bar{b}$
cross-section: the quark initiated subprocess ($q\bar{q}\to b\bar{b}b\bar{b}$) and the gluon initiated
one ($gg\to b\bar{b}b\bar{b}$).
In this contribution we presents results obtained for the quark induced subprocess
Throughout the whole calculation we treat the $b$-quarks as massless, an approximation which is well 
motivated by the LHC kinematics and the applied cuts, and neglect the effect of a heavy top quark altogether. 
Furthermore we neglect the $b$-quark contribution in the parton distribution functions, due to the smallness of
the $b$ PDF.

\subsection{Virtual corrections}
The calculation of the virtual corrections is based on the GOLEM approach
\cite{Binoth:2005ff,Binoth:2008gx,Binoth:1999sp,Binoth:2006hk} for one loop amplitudes.
It uses a Feynman diagrammatic representation generated by QGRAF \cite{Nogueira:1991ex}.
{\tt FORM} \cite{Vermaseren:2000nd}
is used to map the helicity amplitudes to a tensor form factor
representation as defined in \cite{Binoth:2005ff,Reiter:2009kb,Reiter:2009dk}. The latter
is exported to {\tt Fortran95} code such that
it can be linked to the form factor library {\tt golem95} \cite{Binoth:2008uq}.
For the code generation we have developed a dedicated optimisation tool
which can also be used for other purposes \cite{Reiter:2009ts}.
To validate the results a second independent calculation has been performed.
There we did a symbolic tensor reduction down to genuine scalar integrals where we used
 {\tt FeynArts} and {\tt FeynCalc} \cite{Hahn:1998yk} to generate the amplitudes.
The tensor integrals are reduced symbolically to
scalar integrals also using the formalism described in \cite{Binoth:2005ff}. 

More recently we have independently calculated the virtual corrections using the SAMURAI 
package~\cite{Mastrolia:2010nb}. SAMURAI is a package for the reduction of one-loop amplitudes at the 
integrand level based on the OPP approach~\cite{Ossola:2006us,Ossola:2007bb}. Combined with a modified 
version of our matrix element generator the code for the numeric evaluation of the matrix element was 
generated in a fully automated way. 

Using this setup we were able to reduce the CPU time needed for the evaluation of the virtual 
corrections of a single phase space point to around $0.4$s, which is around a factor of $10$ faster 
compared to the previous approach.

This improvement in the code speed is a fundamental step towards the computation of the gluon
induced subprocess which, due to the larger number of Feynman diagrams and their increased complexity,
is computationally more intensive than the quark initiated one.

\subsection{Real emission}
The tree level and real emission matrix elements have been generated using MadGraph \cite{Stelzer:1994ta},
for the subtraction terms we used MadDipole \cite{Frederix:2008hu,Frederix:2010cj} 
which is based on the dipole formalism \cite{Catani:1996vz,Catani:2002hc}.
For the full proton initiated process there are three subprocesses contributing: 
$q \bar{q} \to b \bar{b} b \bar{b} g$, $g g \to b \bar{b} b \bar{b} g$ and $q g \to b \bar{b} b \bar{b} q$. 
Here we restrict ourselves to the first one. 
The subtraction terms and the real emission matrix element have been implemented into the MadEvent framework 
\cite{Maltoni:2002qb}.

A second implementation of the subtraction terms in Whizard \cite{Kilian:2007gr,Moretti:2001zz} has been
used to ensure the correctness of our code.

\section{Results}
\begin{figure}[ht]
\begin{center}
\includegraphics[height=5cm]{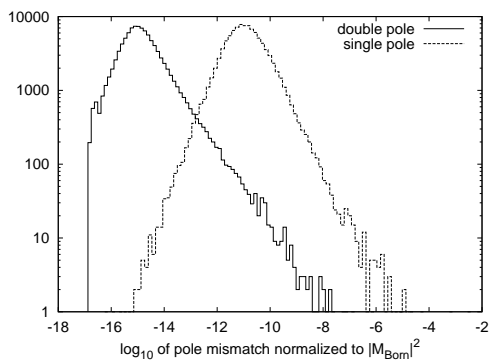}
\end{center}
\caption{Cancellation of the poles from the virtual corrections
and the subtraction terms normalised to the Born. To produce this plot we 
integrated the process $q \bar{q} \to b \bar{b} b \bar{b}$
using $10^5$ phase space points and the cuts described in the text.}
\label{fig:poles}
\end{figure}

In order to provide a check of the numerical stability of our implementation we looked
at the cancellation of the double and single poles between the virtual corrections and dipole 
subtraction terms.
The result is presented in Fig. \ref{fig:poles} and shows that the cancellation happens within 
double precision accuracy for both single and double poles, thus confirming that possible numerical 
instabilities are well under control.

The numerical results we present in this section were obtained for $\sqrt{s} = 14$ TeV center of mass
energy. In a first step we cluster the final state particles to $b$-jets using the $k_T$-algorithm
\cite{Blazey:2000qt}. The resulting jets have to pass the following cuts:
\begin{eqnarray}
 p_{T}(b_j) &>& 30 \;\; \textrm{GeV} \nn \\
|\eta(b_j)| &<& 2.5 \nn\\
\Delta R(b_i, b_j) &>& 0.8  \quad .
\end{eqnarray}
One of the main goals of calculating higher orders corrections is to reduce the
dependence of the result on renormalisation and factorisation scale.
\begin{figure}[ht]
\mbox{} 
\includegraphics[height=4.7cm]{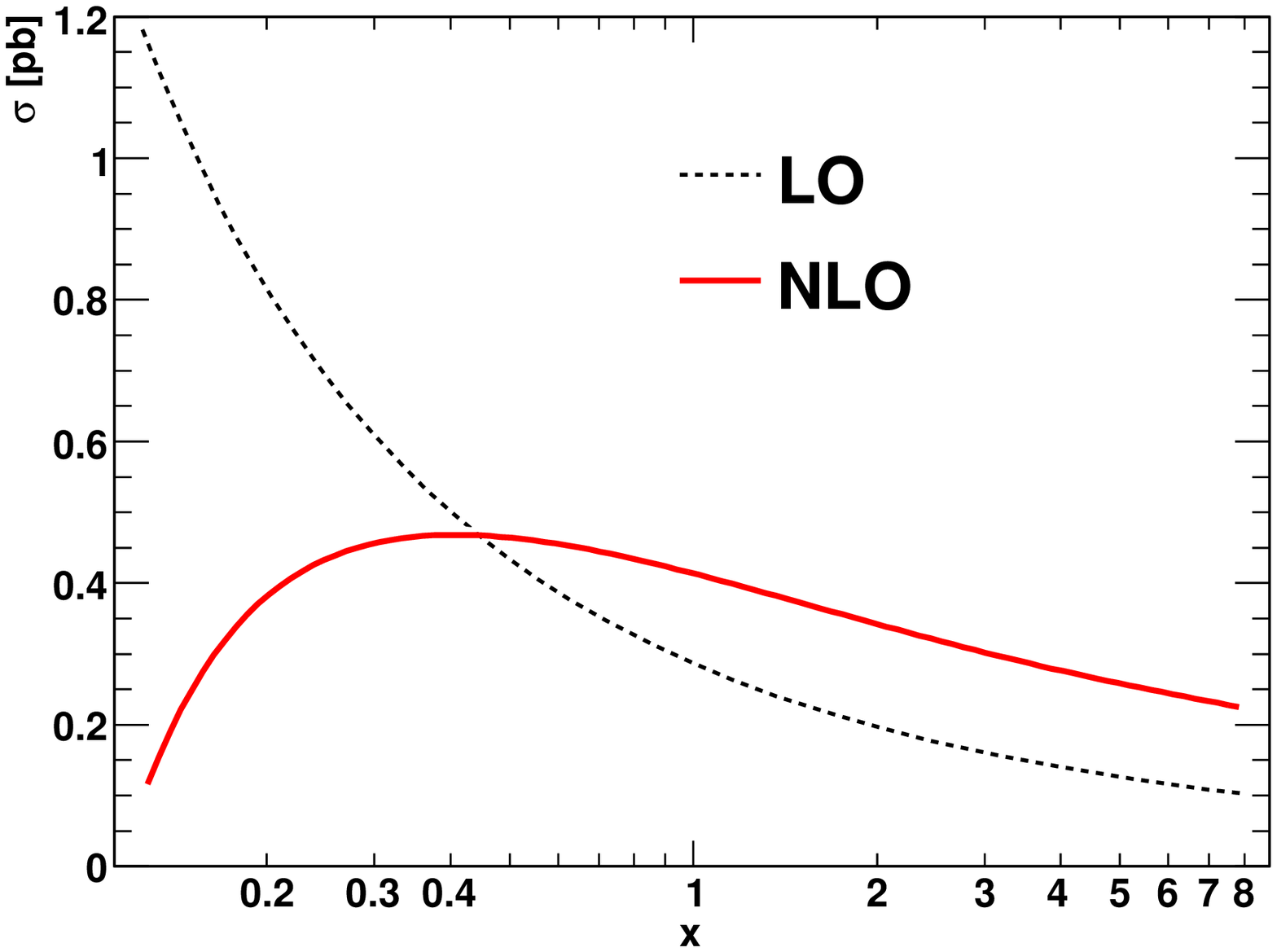} \hspace{0.5cm}
\includegraphics[height=4.7cm]{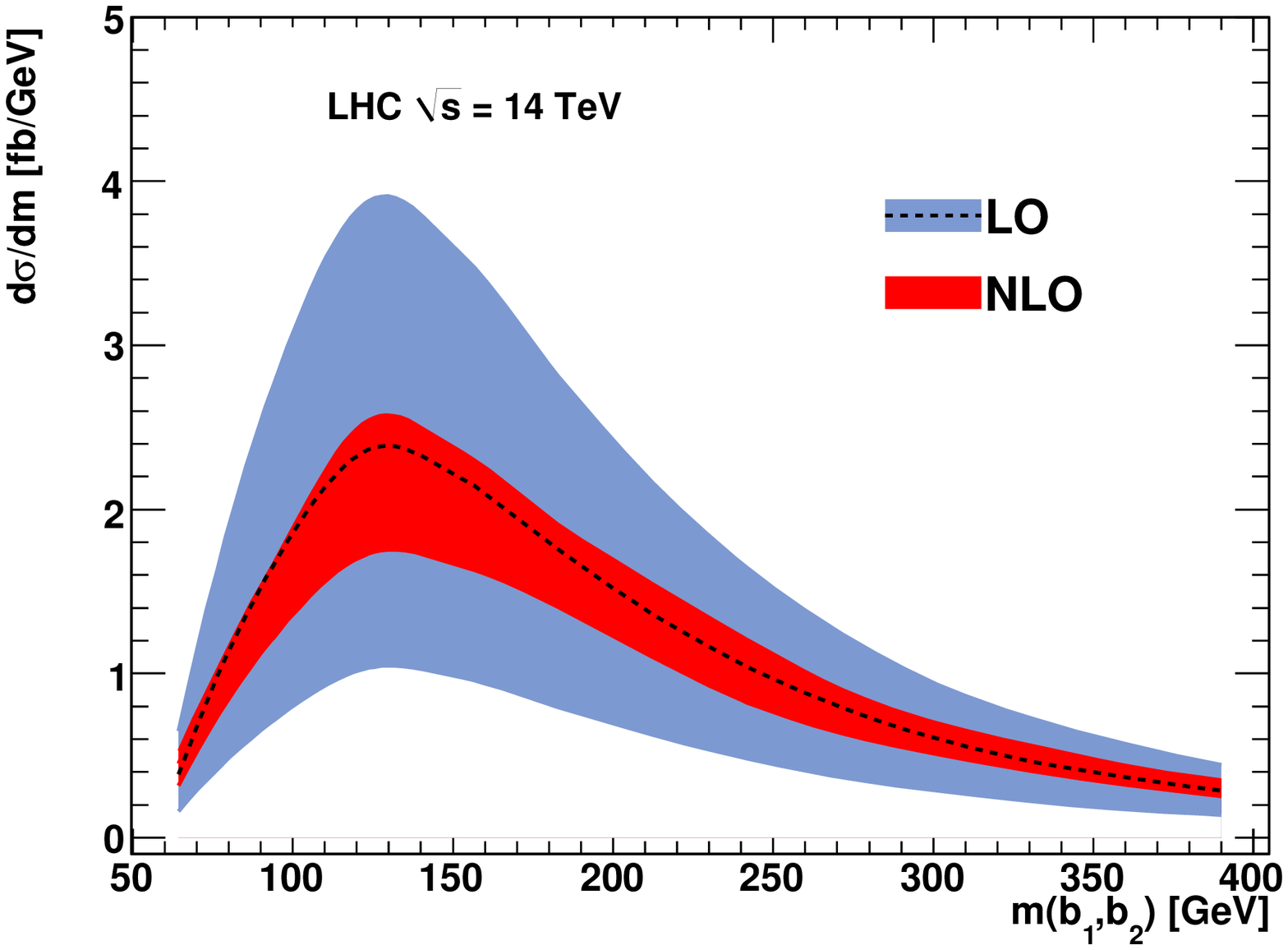}
\caption{Dependence on the renormalisation scale for the total cross section (left plot)
and for the distribution of the invariant mass of the two $b$-jets with the higest $p_T$
(right plot) when varying the scale around $\mu_r= x \cdot \mu_0=x \cdot \sqrt{\sum_j p_T^2(b_j)}$ with
$x \in [\frac{1}{8},8]$ (left plot) and $x \in [\frac{1}{4},2]$. (right plot). The dashed line in the right
plot denotes the leading order prediction for $x=1/2$.}
\label{fig:scale}
\end{figure}
In Fig. \ref{fig:scale} we compare the dependence on the renormalisation scale for the total cross
section and for the invariant mass distribution of the two $b$-jets with the highest $p_T$ at LO and 
NLO.
As expected, a clear reduction on the scale dependence can be observed. The central scale is
chosen to be $\mu_0=  \sqrt{\sum_j p_T^2(b_j)}$ and the scale is then varied with
$\mu_r=x \cdot \mu_0$. As we have not included all possible
initial states we kept the factorisation scale fixed to be $\mu_F = 100$ GeV.

\section{Acknowledgments}
N.G. is supported by the U.~S.~Department of Energy under contract No.~DE-FG02-91ER40677.
T.R. has been supported by the Foundation FOM, project FORM 07PR2556.

\end{document}